\documentstyle[amssymb,twocolumn,epsfig,aps]{revtex}

\input{psfig.sty}

\begin{document}
\draft
\title{All Optical Formation of an Atomic Bose-Einstein Condensate}
\author{ M. D. Barrett, J. A. Sauer, and M. S. Chapman}
\address{School of Physics, Georgia Institute of Technology, \\
Atlanta, Georgia 30332-0430}

\maketitle

\begin{abstract}
We have created a Bose-Einstein condensate of $^{87}\text{Rb}$ atoms
directly in an optical trap. We employ a quasi-electrostatic dipole force
trap formed by two crossed $\text{CO}_{2}$ laser beams. Loading directly
from a sub-doppler laser-cooled cloud of atoms results in initial phase
space densities of $\sim 1/200$. Evaporatively cooling through the BEC
transition is achieved by lowering the power in the trapping beams over $%
\sim 2$ s. The resulting condensates are $F=1$ spinors with $3.5\text{ x }%
10^{4}$ atoms distributed between the $m_{F}=(-1,0,1)$ states.
\end{abstract}



\narrowtext

The first observation of Bose Einstein condensates (BEC) in dilute
atomic vapors in a remarkable series of experiments in 1995
\cite{jilabec,mitbec,hulet96} has stimulated a tremendous volume
of experimental and theoretical work in this field. Condensates
are now routinely created in over 30 laboratories
around the world, and the pace of theoretical progress is equally impressive %
\cite{stringari}. The recipe for forming a BEC is by now well-established %
\cite{cornellreview,ketterlereview}. The atomic vapor is first pre-cooled,
typically by laser cooling techniques, to sub-mK temperatures and then
transferred to a magnetic trap. Further cooling to BEC is then achieved by
evaporatively cooling the atoms in the magnetic trap using energetically
selective spin transitions \cite{evapreview}.

All-optical methods of reaching the BEC phase transition have been pursued
since the early days of laser cooling. Despite many impressive developments
beyond the limits set by Doppler cooling, including polarization gradient
cooling \cite{pgc}, VSCPT \cite{vscpt}, Raman cooling \cite%
{churaman1,weisscooling}, and evaporative cooling in optical dipole force
traps \cite{chu,thomasevap,grimmevap}, the best efforts to-date produce
atomic phase space densities $n\lambda _{dB}^{3}$ a factor of 30 away from
the BEC transition \cite{weisscooling}. The principal roadblocks have been
attributed to density-dependent heating and losses in laser cooling
techniques, residual heating in optical dipole force traps or the
unfavorable starting conditions for evaporative cooling. Hence, optical
traps have played only an ancillary role in BEC experiments. The MIT group
used a magnetic trap with an `optical dimple' to reversibly condense a
magnetically confined cloud of atoms evaporatively cooled to just above the
phase transition \cite{mitreversible}. Additionally, Bose condensates
created in magnetic traps have been successfully transferred to shallow
optical traps for further study \cite{mitoptical,mitspinor,kasevich}. In all
these cases, however, magnetic traps provided the principle increase of
phase space density (by factors up to $\sim 10^{6}$) to the BEC transition.

In this letter, we present an experiment in which we have created
a Bose condensate of $^{87}$Rb atoms directly in an optical trap
formed by tightly focused laser beams. Following initial loading
from a laser cooled gas, evaporative cooling through the BEC
transition is achieved by simply lowering the depth of the optical
trap. Our success is due in part to a high initial phase space
density realized in the loading of our optical dipole trap and in
part to the tight confinement of the atoms that permits rapid
evaporative cooling to the BEC transition in $\sim 2$ s. This fast
evaporation relaxes considerably the requirement for extremely
long trap lifetimes typical of magnetic trap BEC experiments.
Additionally, in contrast to magnetic traps which only confine one
or two magnetic spin projections \cite{jilaspinor}, our technique
is spin-independent and the condensates that we form are $F=1$
three-component spinors \cite{mitspinor}.

We utilize a crossed-beam optical dipole force trap employing tightly
focused high-powered (12 W) infrared ($\lambda =10.6$ $\mu $m) laser beams.
For trapping fields at this wavelength, the trapping potential for the
ground state atoms is very well approximated by $U({\bf r})=-\frac{1}{2}%
\alpha _{g}\left| E\left( {\bf r}\right) \right| ^{2}$ where $\alpha _{g}$
is the ground state dc-polarizability of the atom (5.3 x 10$^{-39}$ m$^{2}$C$%
/$V for atomic rubidium) and $E\left( {\bf r}\right) $ is the electric field
amplitude. A significant feature of these `quasielectrostatic' traps (QUESTs) %
\cite{knize} is that heating due to spontaneous emission of the
atoms is completely negligible. In contrast to previous QUESTs
that have employed single focused beams
\cite{knize,thomasevap,grimmevap} or standing wave configurations
\cite{hansch}, we employ here a cross-beam geometry \cite{chu} to
provide a balance of tight confinement in three dimensions ($\sim
1.5$ kHz oscillation frequencies at full power) and a relatively
large loading volume.

Our experiments begin with a standard vapor loaded MOT. The trapping beams
consist of three orthogonal retroreflected beams in the $\sigma^{+}-\sigma
^{-}$ configuration. They are tuned 15 MHz below the 5S$_{1/2}$-5P$_{3/2}$ $%
F=2\rightarrow F^{\prime }=3$ transition of $^{87}$Rb and each of the three
beams has a waist of 0.7 cm and a power of 25 mW. A repump beam tuned to the
$F=1\rightarrow F^{\prime }=2$ transition is overlapped with one of the MOT
trapping beams. The MOT is loaded for 5 s directly from the thermal vapor
during which we collect 30 x 10$^{6}$ atoms. After loading the MOT, the trap
configuration is changed to maximize the transfer of atoms to the optical
trap. The repump intensity is first lowered to $\sim $10 $\mu $W/cm$^{2}$
for a duration of 20 ms, and then the MOT trap beams are shifted to the red
by 140 MHz for a duration of 40 ms. At this point the MOT beams are
extinguished and the current in the MOT coils is turned off. In order to
optically pump the atoms into the $F=1$ hyperfine states, the repump light
is shuttered off 1 ms before the trap beams are extinguished; we measure the
efficiency of the optical pumping to the $F=1$ state to be $>95\%$. The CO$%
_{2}$ laser beams are left on at full power (12 W) throughout the MOT
loading and dipole trap loading process.

The trapping beams are generated from a commercial CO$_{2}$ gas laser ($%
\lambda =10.6$ $\mu $m). The beams are tightly focused and intersected at
right angles; one beam is oriented in the horizontal direction and one beam
is inclined at 45$^{\text{o}}$ from the vertical direction. Each beam passes
through an acousto-optic modulator to provide independent control of the
power in the two beams. Additionally, the beams are frequency shifted 80 MHz
relative to each other so that any spatial interference patterns between the
two beams are time-averaged to zero \cite{weiss}. Each beam has a maximum
power of 12 W, and the beams are focused to a minimum waist $\lesssim $ 50 $%
\mu $m with $f=38$ mm focal length ZeZn aspherical lenses inside the chamber.

Following standard techniques, the number of trapped atoms and
their momentum distribution are observed using absorptive imaging
of the released atoms. The trapped atoms are released by suddenly
($<$ 1 $\mu $s) switching off the trapping laser beams. Following
a variable ballistic expansion time (typically 2-20 ms), the cloud
is illuminated with a 50 $\mu $s pulse of $F=1\rightarrow
F^{\prime }=2$ light applied to the atoms concurrent with a
vertically oriented circularly polarized probe beam tuned to the
$F=2\rightarrow F^{\prime }=3$ transition. The probe intensity is
$\sim $0.3 mW/cm$^{2}$, and each atom scatters up to 150 photons
from the probe with no observable blurring of the cloud. The
shadow of the atom cloud is imaged onto a slow-scan CCD camera.
The measured spatial resolution of our imaging system is $\lesssim
10$ $\mu $m. From these images, the number of atoms and their
temperature are determined. Together with the trap oscillation
frequencies (which are measured directly using parametric
excitation) the spatial density, elastic collision rate, and phase
space density can be derived.

We first discuss the properties of the trap following loading from the MOT
without employing forced evaporative cooling. The earliest that we observe
the trapped atoms is 100 ms after loading to allow for the untrapped atoms
to fall away. At this time, the trap contains 2 x 10$^{6}$ atoms at a
temperature of 75 $\mu $K. Maintaining full power in the trap beams, we
observe a rapid evaporation from the trap in the first 1.5 s, during which
of two-thirds of the trapped atoms are lost and the temperature falls to 38 $%
\mu $K. The relative phase space density increases by a factor of 3 during
this stage. This rapid evaporation gives way to a much slower exponential
trap decay with a time constant of 6(1) s. The temperature continues to fall
gradually to a final value of 22 $\mu $K on a 10 s timescale.

The mean frequency of the trap at these powers is measured to be
1.5(2) kHz, from which we infer an initial peak density of the trap of $\sim $2 x 10%
$^{14}$ atoms$/$cm$^{-3}.$ The initial phase space density
following the rapid free evaporation stage is calculated to be
1/200 (or 1/600 assuming equal distribution of the atoms in the
three trapped internal states). Using the measured value for the
three body loss rate \cite{jila3body} yields an initial 3-body
loss rate constant of 1.3 s$^{-1}$. Hence it is possible that
3-body inelastic collisions contribute to the rapid initial loss
of atoms, however we note that such collisions do not explain the
observed cooling. The estimated elastic collision rate at this
point is a remarkable 12 x 10$^{3}$ s$^{-1}$---higher than the
trap oscillation frequency. Although the derived quantities must
be taken with some caution, we note that densities of 3 x
10$^{13}$ atoms$/$cm$^{-3}$ and a phase space density of 1/300
have been reported in a 1-D CO$_{2}$ lattice trap with $^{85}$Rb
(albeit with much fewer atoms) \cite{hanschresolv}, and we have
measured similar results in our laboratory in 1-D CO$_{2}$ lattice
traps with $^{87}$Rb. It is not clear what yields these extremely
high initial densities in these traps at these low temperatures. A
blue detuned
Sisyphus process is suggested in \cite{hanschresolv} (densities of $\sim $10$%
^{13}$ atoms$/$cm$^{-3}$ have been observed in a dipole trap with
blue detuned Sisyphus cooling \cite{saloman}), but we suspect that
there is an effective dark SPOT \cite{darkSPOT} in the repump beam
at the trap center induced by the AC stark shift due to the dipole
trap beams, which reduces density-limiting interactions with the
MOT light fields. Using our trap parameters, we estimate that the
repump scattering rate is reduced at the trap center by a factor
of 40 relative to outside the trap. Our improved performance even
relative to the 1-D lattice traps may be due to more efficient
loading from the `tails' of the crossed trap geometry. In any
case, the net result for our trap is a very high initial phase
space density combined with a large number of atoms and a fast
thermalization time---together these conditions provide a very
favorable starting point for evaporative cooling.

Efficient forced evaporation requires selectively ejecting the more
energetic atoms from the trap such the the remaining atoms rethermalize at a
lower temperature with a higher net phase space density. For optical traps,
the simplest way to force evaporation is to lower the trap depth by
decreasing the power in the trap beams. This technique was used in one of
the first demonstrations of evaporative cooling in alkali atoms \cite{chu},
where, starting with only 5000 atoms, a phase space density increase of $%
\sim 30$ was realized. The drawback to this method is that lowering the trap
depth also lowers the trap oscillation frequency and the rethermalization
rate. Hence the evaporation rate can slow down prohibitively. In our case,
although the rethermalization rate falls by a factor of 50 by the end of the
evaporation cycle, it nonetheless remains fast enough to allow us to reach
the BEC transition in 2.5 s as described below.

Our procedure for forced evaporation is as follows: immediately after
loading the trap, the power in both trap beams is ramped to 1 W in 1 s. The
power in both beams is then ramped to a variable final power in 1 s and
maintained at this low level for 0.5 s, after which the atoms are released
from the trap and imaged as discussed above. Fig. 1 shows three such images
at different final powers for the trap lasers. The left image shows a cloud
well above the BEC transition point, and reveals an isotropic gaussian
momentum distribution expected for a thermal cloud of atoms in equilibrium.
As the evaporative cooling proceeds to lower powers, a bimodal momentum
distribution appears with a central non-isotropic component characteristic
of Bose condensates (Fig. 1, center). As the power is lowered further, the
central, non-spherical peak of the distribution becomes more prominent, and
the spherical pedestal diminishes. At a trap power of 190 mW (Fig. 1,
right), the resulting cloud is almost a pure condensate and contains 3.5 x 10%
$^{4}$ atoms. For lower powers, the cloud rapidly diminishes as the trap can
no longer support gravity. Line profiles of the images in Fig. 1 are shown
in Fig. 2 for quantitative comparison. The profiles are taken along the
orientation of the minor axis of the condensate. Also shown are gaussian
fits to the data in the wings that clearly show the bimodal nature of the
momentum distributions near the BEC phase transition.

The critical condensation temperature $T_{c}$ is a function of the
number of atoms $N$ and the mean trap frequencies
$\bar{\omega}=(\omega_1 \omega_2 \omega_3)^{1/3}$ according to
$k_{B}T_{c}=\hbar \bar{\omega}(N/1.202)^{1/3}.$ Below the critical
temperature, the condensate fraction should grow as $N_{0}/N=1-(T/T_{c})^{3}$%
\cite{cornellreview,ketterlereview}. We obtain the condensate
fraction and the temperature of the normal component from 2D fits
to the absorptive images using the methods described in
\cite{ketterlereview}. In Fig. 3, the condensate fraction is
plotted versus normalized temperature, $T/T_{c}(N,\bar{\omega})$
near the critical point, where we assume that $\bar{\omega}^{2}$
is proportional to the trap power. The agreement with the
theoretical curve is reasonable given the scatter in the data. At
the trap power of 350 mW, which is near the critical point, the
trap contains 180,000 atoms at a temperature of 375 nK. Together
with the measured trap frequencies at this power $\omega
_{1},\omega _{2},\omega _{3}=2\pi (72,175,350)$ Hz, we infer a
phase space density at this point of 1.4 (or 0.45 if the three
internal states are equally populated), a density of 4.8 x
$10^{13}$ cm$^{-3}$ and an elastic collision rate of 300 s$^{-1}$.

We have measured the growth of the freely expanding condensate for
expansion times from 5-20 ms. The measured aspect ratio of the
cloud is 1.5(0.1) and shows a slight increase for longer expansion
times. The measured size of the minor axis of the cloud grows
linearly with time (within the limits of our imaging resolution)
to a size of 100 $\mu $m (full-width) at 15 ms. The observed 2D
projection of the cloud is oriented at a $30^{\text{o}}$ angle
relative to the horizontal trap. Quantitative comparison of the
observed condensates with theory requires knowledge of the trap
oscillation frequencies and as well as orientation of the
principal axes of the trap. Although we can measure the
frequencies, the principal axes depend on the beam ellipticity and
alignment. In our case, the trap beams are aberrated due to
off-center propagation through the focussing lenses (required to
overlap the beams), and the fact that the laser output beam is not
a pure gaussian TEM$_{00}$ spatial mode. Nonetheless, using the
measured trap frequencies, scaled for power, mean field theory
(see e.g. \cite{stringari}) predicts aspect ratios of 1.2-1.7 for
a 15 ms expansion and cloud sizes of 60-130 $\mu $m, consistent
with our observations.

The $1/e$ lifetime of the condensate is measured to be 3.5(1) s, which is
somewhat smaller than the 6(1) s lifetime observed in the trap. We observe
no residual heating of the condensate for the lifetime of the condensate,
although because the trap potential is quite shallow, it is possible that
any heating would be offset by subsequent evaporation from the trap.

The trapped atoms are optically pumped during loading into the
$F=1$ groundstate, however, no attempt is made to further
optically pump the atoms into a single $m_{F}$ state. Hence we
expect the population to be a mixture of the $m_{F}=-1,0,1$ spin
projections. To measure the spin content of our condensate, we
apply a weak field gradient to the cloud after the cloud is
released from the dipole trap \cite{mitspinor}. An absorptive
image of the cloud is shown in Fig. 4, and reveals that the
condensate is comprised of a mixture of spin-states. Each cloud
has a similar non-spherical momentum distribution characteristic
of condensates. Determining the relative weights of the spin
states is complicated by the differing matrix elements of the
probe transitions. Nonetheless, if we assume that the atoms are
all optically pumped by the circularly polarized probe and that
all atoms
scatter the same number of photons, we infer a 3:1:1 weighting of the $%
m_{F}=-1,0,1$ mixture respectively for the image shown in Fig. 4.

In summary, we have realized a Bose condensate of $^{87}\text{Rb}$
atoms directly in an optical trap.  This technique seems promising
for creating and studying more complex spinor condensates (e.g.
$^{85}\text{Rb},F=2$) as well as multi-atom mixtures.
Additionally, by eliminating the need for a magnetic trap, it may
be possible to realize condensates in atoms or molecules lacking a
suitable magnetic moment, perhaps by using sympathetic cooling to
precool the sample. Finally, our technique offers considerable
experimental simplicity and speed, easing the requirement for
ultra high vacuum environments, and eliminating the need for
strong magnetic trapping fields. \qquad

We would like to acknowledge the technical assistance of D. Zhu and helpful
discussions with T.A.B. Kennedy, C. Raman, S. Yi, and L. You. This work was
supported by the National Security Agency (NSA) and Advanced Research and
Development Activity (ARDA) under Army Research Office (ARO) contract number
DAA55-98-1-0370.

\begin{figure}[tbp]
\centerline{\epsfig{figure=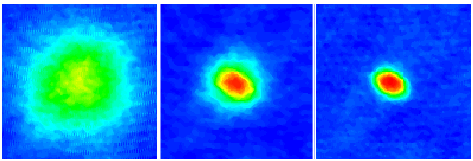,width=3.0in}}
\caption{Absorptive images (false color) of atomic cloud after 10
ms free expansion for different final trap laser powers. (a)
thermal cloud above BEC transition (P = 480 mW), (b) thermal -
condensate mixture (P = 260 mW) (c) pure condensate (P = 190 mW).
Field of view is 350 $\protect\mu$m. }
\label{fig1}
\end{figure}

\begin{figure}[tbp]
\centerline{\epsfig{figure=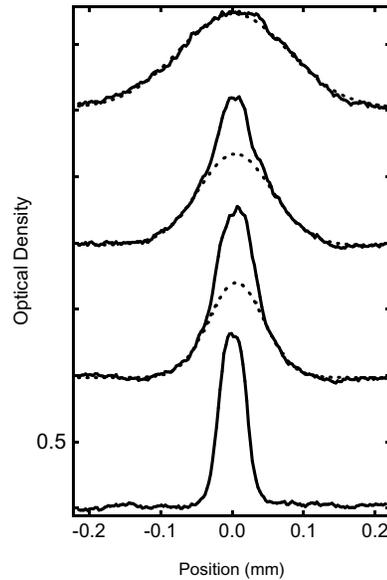,width=2.0in}}
\caption{One-dimensional profiles through the images of Fig 1.
(solid) along with gaussian fits to the wings of the profile
(dashed). Trap laser powers 480, 310, 260, 190 mW, top to bottom
respectively.} \label{fig2}
\end{figure}

\begin{figure}[tbp]
\centerline{\epsfig{figure=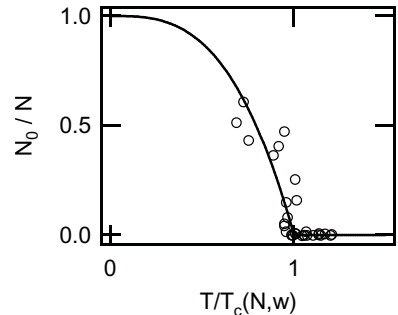,width=2.0in}}
\caption{(circles) Condensate fraction (ratio of condensate atoms
to total number of atoms) vs. scaled temperature,
$T/T_{c}(N,\bar{\omega})$. Also shown is the theoretical
prediction (solid curve).} \label{fig3}
\end{figure}

\begin{figure}[tbp]
\centerline{\epsfig{figure=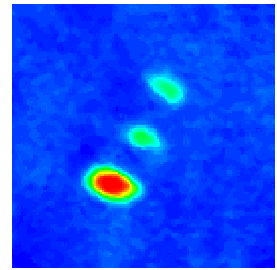,width=1.5in}}
\caption{Absorptive image of atomic cloud after 10 ms free
expansion in a Stern-Gerlach magnetic field gradient. Three
distinct components are observed corresponding to $F=1,$
$m_{F}=(-1,0,1)$ spin projections from bottom to top,
respectively. } \label{fig4}
\end{figure}

\end{document}